\documentclass[12pt]{article}

\usepackage{epsfig}
\begin{document}
\centerline{\Large Einstein's Theory of Gravity and the Problem of Missing Mass}
\vspace{5mm}
 \centerline{ Pedro G. Ferreira}
 \centerline{ Astrophysics, University of Oxford, Keble Road, Oxford, OX1 3RH, U.K.}
\centerline {  {\tt p.ferreira1@physics.ox.ac.uk} }
\vspace{3mm}
 \centerline{Glenn D. Starkman}
 \centerline {CERCA/ISO/Physics, Case Western Reserve University, Cleveland,Ohio,USA}
\centerline {  {\tt glenn.starkman@case.edu}  }

\vspace*{10mm}
\noindent
{\bf The observed matter in the universe accounts for just 5$\%$ of the observed gravity.
A possible explanation is that Newton's and Einstein's theories of gravity fail where gravity is either weak or enhanced.
The modified theory of Newtonian dynamics (MOND) reproduces, without dark matter, spiral-galaxy orbital motions
and the relation between luminosity and rotation in galaxies,
although not in clusters.
Recent extensions of Einstein's theory are theoretically more complete. 
They inevitably include dark fields that seed structure growth, 
and they may explain recent weak lensing data.   
However, the presence of dark fields reduces calculability and comes at the expense of 
 the original MOND premise -- 
that the matter we see is the sole source of gravity.
Observational tests of the relic radiation, weak lensing,  
and the growth of structure may distinguish modified gravity from dark matter.
}

\section*{Introduction}
The problem of missing mass has been with us more than 70 years: Given the amount of directly observable matter,
general relativity (GR, Einstein's theory of gravity) produces too  little gravity to account for a host of
observations.  On scales of one to tens of kiloparsecs, the observed random or coherent velocities of stars and gas
are much greater than the escape velocity in the self-gravity of those same stars, gas and dust. 
The same is true for galaxy clusters on much larger scales. 
Gravitational potentials around galaxy clusters, deeper than would be produced by the observed matter,
are also needed to explain observed gravitational lensing; 
that is. the deformation of light bundles from background galaxies.

Evidence also exists for anomalously strong gravity  on the largest observable scale: 
out to the cosmological horizon.  
In a universe that contained only ordinary matter (often called baryonic matter, encompassing protons and neutrons, which make up over 99.9\% of the mass
of ordinary matter), the growth of structures, such as galaxies and clusters of galaxies, would be suppressed.  During recombination, when that universe was approximately $400,000$ years old, 
 the seeds for galaxies and clusters would be entirely erased by dissipation 
(known as Silk damping) and no structure would form on scales up to many tens of megaparsecs.

The now standard solution of these dynamical mysteries is that
an unobserved form of mass, which exceeds the observed mass of both galaxies
and clusters,  provides the gravity that prevents them from flying apart, 
increases lensing and prevents Silk damping.
The missing mass neither shines nor absorbs or scatters light enough to be directly detected by our telescopes. It should be close to pressureless and must be non relativistic well before recombination. It is therefore known as {\it cold dark matter} (CDM).
A number of candidate particles have been proposed that have these properties 
and their  behaviour in the Universe has been studied in exquisite detail.  

\begin{figure*}[htb]
\center
\epsfig{figure=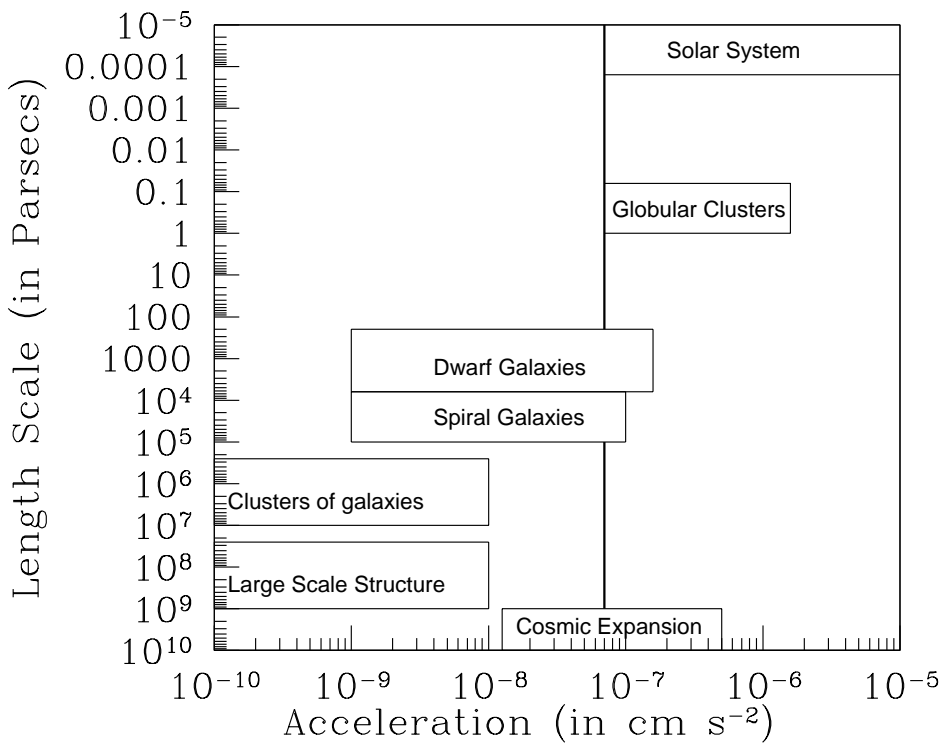,width=13cm}
\vspace{-30pt}
\caption{ Evidence for dark matter or for deviations from GR tend to appear in systems
in which the acceleration scale is weak (to the left of the  solid horizontal line) at about $7\times10^{-8}$ cm $^{-2}$.
There is strong evidence for either of the above in dwarf galaxies, spiral galaxies, clusters of galaxies,
the large scale structure of the Universe and in the expansion of the Universe itself.
}\label{fig1}
\end{figure*}

An altogether different approach can be taken if one notes that the evidence for missing mass
arises because of a mismatch between the gravitational field one would {\it predict} from the observed
mass distribution in the Universe and the {\it observed} gravitational field.  The observed discrepancies
arise when the effective gravitational acceleration is around, or below, 
$a_0\simeq10^{-8}$ cm sec$^{-2}$; that is in a regime of
very weak gravitational field. 
Perhaps the Newtonian theory of gravity- and GR-  break down in this regime. In this Review we provide an
updated assessment of this theory.

\section*{Modifying Newtonian Gravity}

The possibility that Newtonian gravity and GR do not accurately describe very weak gravitational
fields was proposed
more than 25 years ago. Milgrom suggested that Newton's second law, $ {\vec F}=m{\vec a} $ (where ${\vec F}$ is the
gravitational force applied to a unit of mass $m$ to produce an acceleration ${\vec a}$) is modified when  
gravity is weak, to ${\vec F}=m\left(\vert{\vec a}\vert/a_0\right){\vec a}$ \cite{Milgrom1983}. 
 This proposal has been named modified Newtonian dynamics (MOND).
More modern versions of MOND cast it instead as a modified theory of gravity,
altering the Newton-Poisson equation that relates the gravitational
force to the distribution of mass density $\rho$ that is responsible for it \cite{Bekenstein1984}.

MOND has a number of appealing features. It independently explains the empirical Tully-Fisher relation
between the luminosity, ${\cal L}$, of a spiral galaxy
and its asymptotic rotational velocity, $v$: ${\cal L}\propto v^4$ where $G$ is the gravitational
constant and $M$ is the baryonic mass of the galaxy.     Given the ratio of the baryonic mass of
these spirals to their luminosity (the mass-to-light ratio), this is equivalent to
 $(a_0G)M=v^4$, exactly what would be predicted by MOND.
 A systematic study of a wide range of spiral galaxies pins 
 the acceleration scale to be unique:
$a_0\simeq 1.2\times10^{-8}$ cm s$^{-2}$.
Furthermore, the detailed features of the rotational velocity 
as a function  of radius are predicted by the baryonic mass distribution \cite{McGaugh1998}.
MOND has also been used to predict the analogue of the Tully-Fisher
relation for elliptical galaxies (the Faber-Jackson relation
between baryonic mass and velocity dispersion), the existence of galaxies with low surface brightness, 
and an upper limit on the mean surface brightness of spiral galaxies (known as Freeman's law) \cite{Sanders2002}.

\begin{figure*}[htb]
\center
\epsfig{figure=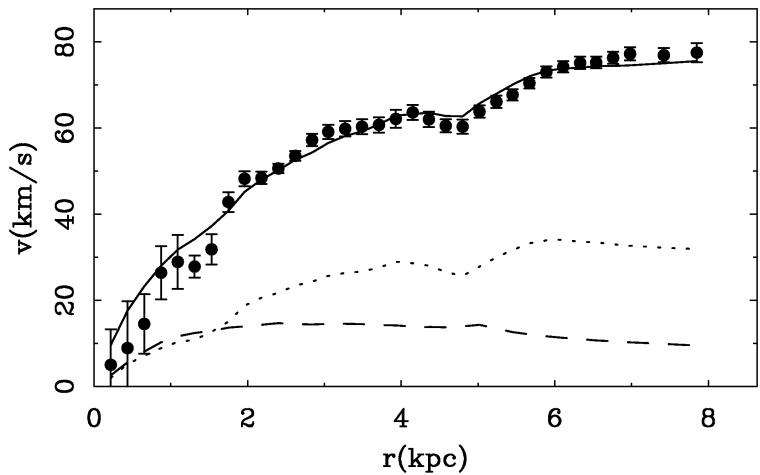,width=10cm}
\vspace{-10pt}
\caption{ The MOND rotation curve of the galaxy NGC1650  (solid line) reproduces observed features
with surprising fidelity given just one free parameter -- the mass-to-light
ratio $M/L=0.43$.  Also shown is the Newtonian rotation curve that would
result from  just the gas (dotted) or just the stars (dashed). The quality of the rotation curve fit
from MOND is generic.
(Figure courtesy of Mcaugh; originally published in \cite{Begeman1991})
}\label{fig2}
\end{figure*}
Attempts to resolve the mass discrepancy on the scale of clusters of galaxies have been more problematic.
One is obliged either to use a value of $a_0$ that is different from the one used for galaxies 
or to assume the existence of a small amount of dark matter. 
If one of the three types of neutrino (electron, muon or tau)
has a mass of a few electrons volts it would be an ideal candidate
for cluster dark matter in MOND \cite{Sanders2003}. 

An added complication is that gravitating systems cannot be studied in isolation and the external 
gravitational field can play a role in the internal dynamics of disparate objects such as star clusters, molecular clouds and galaxies.  
MOND does not satisfy Birkhoff's theorem (the analogue in gravity of Gauss' law
in electromagnetism)  for real masses  \cite{Lue2004,Dai2008,Wu2007} and this means that the  acceleration of any real probe -- a star
or a cloud of gas -- even if it is located in a spherically symmetric system, depends not just on the mass that is interior to the
probe but on the mass that is exterior as well.  

MOND was developed as a phenomenological
description of spherically (or at least axially) symmetric, non-relativistic, low acceleration systems.  
Until a fully dynamical relativistic theory can be constructed, MOND itself cannot reliably predict, among other things: anomalous accelerations in galaxy clusters, the effects of gravitational lensing of light, the expansion of the Universe  or the growth of structure.

\section*{Relativistic theories of modified gravity}
Despite MOND's successes and failures, for it to be seriously judged as a candidate explanation of anything,
it must be  embedded in a modification of Einstein's GR theory.
 Einstein cast gravity as a geometric theory of spacetime (the combination of space and time). 
The properties of spacetime are encoded in a 4 by 4 symmetric matrix with $10$ free components, 
which is called the metric, ${\bf g}$, which is itself a function of space and time.  From this metric, one can construct various geometric quantities, such as the overall curvature of spacetime, $R$, known as the Ricci scalar, 
the Ricci tensor ${\bf R}$ and the Riemann tensor, ${\bf {\cal R}}$. 

Einstein proposed that the energy content of the Universe would source these various quantities, 
curving spacetime according to a fixed set of rules, called the Einstein field equations.
The different components of the Universe would in turn respond to the curvature of spacetime: 
In the absence of non gravitational forces, they would follow geodesic paths that could be derived from the metric. 

Einstein's theory must be tampered with to incorporate MOND. 
There are two possible ways of modifying it. 
One way is to change how curvature responds to the presence of matter. 
The rules that Einstein posited for deriving gravity start with the simplest
of all the quantities that encode curvature, the Ricci scalar. A first step away from Einstein is to bring in other functions of the metric such as the Ricci and Riemann tensors, 
as well as more complicated functions of the Ricci scalar. 
Indeed,  because of the effects of quantum mechanics on the spacetime metric, one {\it expects} the full theory of gravity 
to be a more complex combination of various geometric quantities \cite{Sakharov1968}. 
As a result, one inevitably introduces new gravitational degrees of freedom.

Some of these modifications have consequences  at short distance scales 
and  result in small (although potentially measurable) corrections to standard physics, 
which are insufficient  to reproduce galactic rotation curves.
Others result in modifications to cosmology, but not in MOND-like behavior on galactic scales.
Nevertheless a few proposals have been advanced for modifications of gravity that can play a role on galactic and supra-galactic scales. In \cite{Mannheim1989} the Ricci tensor is replaced by the Weyl curvature and a scalar field
is introduced to play the role of a variable Newton's constant. In \cite{Navarro2006},  a logarithm of the Ricci curvature
is considered as the fundamental action. For these cases and others,
there is an extensive program to explore the theoretical and phenomenological consequences.

A complementary approach is to postulate that light and matter respond to the geometry of space and time differently
than predicted by Einstein. 
The simplest way to implement this is to distinguish the {\it geometric} metric which responds dynamically to the contents of the Universe 
from the {\it geodesic} metric which dictates (in the absence of other, non-gravitational, forces) how those contents propagate through spacetime. 
The simplest such theory relates the two metrics by a location-dependent change of scale, 
known as a {\it conformal} transformation, and endows the scalar field 
describing this transformation with a dynamics of its own. 
This theory has been extensively studied in many contexts and is highly constrained \cite{Fujii2003}. Whereas such a
scalar field
can affect the dynamics of massive bodies, it doesn't modify the propagation of light rays and therefore
will not play a role in phenomena such as gravitational lensing.

More general transformations involve introducing not only a change in scale between the metrics, but  also a distortion of  angles  and this can be done, for example, by introducing a preferred time direction- or a preferred
rest frame. The most natural implementation is to add in a spacetime vector field that has a non-zero value at each point in spacetime; 
in other words to point in some direction in spacetime.  If that direction is chosen to be (on average) the direction that defines the future (
forward in time) as opposed to some direction in space, then the preferred direction will be established  \cite{Sanders1997}. 

Bekenstein \cite{Bekenstein2005} recently proposed a fully relativistic theory that included all of these elements: 
a disformal relation between the geometric  and geodesic metrics, a preferred frame, and modified dynamics for the geometric metric.
For an appropriate choice of an arbitrary but  universal function,
his theory could lead to MONDian dynamics on galactic scales. 
Bekenstein's theory is known as TeVeS, where the T stands for tensor 
(representing the metric), V for the time-like vector field, and S for  the scalar field responsible
for the scale transformation.  
TeVeS  is actually part of a wider class of models that reproduce MOND on 
astrophysical scales.  An alternative subclass of these  models dubbed generalized Einstein-Aether (GEA) theories, include only the time-like vector field and no scalar field \cite{Zlosnik2007}.  

These modifications of Einstein GR are sufficiently well defined that it is possible to make firm
predictions within each model for what should be observed on various astrophysical and cosmological scales. 
Such theories have interesting properties, but whereas Einstein's original proposal is
highly constrained, these more complex proposals are less so.
Inevitably, they  involve extra fields that may come to behave very much like dark matter.

\section*{Observational tests and limitations}

With a relativistic theory of modified gravity in hand, it is possible to make a number of predictions
on a range of scales. For a start, one can focus on the effect that TeVeS or GEA will have on the gravitational field of
compact objects, such as stars or black holes. It has been shown that the atomic spectral lines from the surface of stars
will be affected by TeVeS parameters \cite{DeDeo2003} whereas farther out, there may be directly
detectable preferred frame effects that will modify the Newtonian orbits of nearby objects \cite{Sagi2009}. 
On even larger scales, it has been proposed that the difference in flight time between gravity waves
and neutrinos from, for example, a supernova can be used as a signature of modified gravity theories \cite{Kahya2007}.
As yet, an analysis of millisecond binary pulsars, one of the GR laboratories {\it par excellence}, is still
lacking.

Relativistic theories of modified gravity make specific predictions about the dynamics of the Universe.  
The TeVeS theory has a particular property: The energy density in the extra
fields is always proportional to the energy density of whatever is the dominant form. 
Furthermore, the constant of proportionality is independent of the initial conditions 
and  dependent solely on the fundamental constants of the theory
\cite{Skordis2005}.   
These features lead to a tight constraint on the overall energy density in the extra fields  -- 
it cannot be more than one fifth of the contribution from baryons. 
Thus, unlike dark matter, this energy density is subdominant to the baryonic mass and does not affect the overall expansion rate.
Such behavior can be found in other proposals for modified gravity but it is not generic. It
is challenging to find a parametrization replacing CDM \cite{Li2008},
but such theories  can predict a wide range of cosmological behavior, 
from the highly regular to the unstable, leading to accelerated expansion or to contraction on a finite time scale 
\cite{Bourliot2007,Zlosnik2008}.

Much of the recent advance in cosmology has been acomplished through understanding and measuring
the statistical properties of the growth and morphology of large scale structure, through the cosmic microwave background (CMB) and through surveys of galaxies. 
With relativistic theories of modified gravity it is now possible to make predictions on the largest scales, 
and this has been done for a selection of the
currently proposed models, in particular for the original TeVeS model and for GEA theories. 

Inhomogeneities evolve in a more complex way in these theories than in the case of GR,
with two main new qualitative features. 
First, the extra degrees of freedom drive the initial growth of perturbations; 
they seem to be a necessary piece of the theory and there seems to be no other way to seed structure given the constraints from observations of the CMB
on the amplitude of fluctuations in the baryonic matter density when the universe was 1000
times smaller than it is today.   Gravity alone,  even (stronger) MONDian gravity, appears to be incapable
of growing structure without seeds of such structure that are less coupled to the photons than are baryons \cite{Lue2004}.
TeVeS and GEA avert this conundrum by allowing modes of the new gravitational fields to grow
and seed baryonic structure.  Effectively, these new degrees of freedom act as dark fields \cite{Skordis2005,Dodelson2006}.

A preliminary comparison between the TeVeS theory and both the CMB and large scale structure data indicates
that they are roughly compatible\cite{Skordis2005}. There are a few caveats. First, it may be necessary to include a non-negligible
amount of massive neutrinos with a mass of a few electron volts. This is also the
mass range required by MOND to agree with clusters. It is
still unclear whether this is generic \cite{Ferreira2008}, but if indeed it is, it may be testable in the near future. 
Experiments such as KATRIN (the Karlsruhe tritium neutrino experiment) will bring constraints on the mass of the neutrinos to below 1 electron volt
\cite{KATRIN}.

Second, there is a subtle effect that can emerge on the largest scales. 
In GR, when most of the matter is non-relativistic (in the form of atoms or dark matter), 
perturbations in the metric can be described in terms of one function,
which on small scales is the Newtonian potential that gives rise to the inverse square law of gravity.
In modified theories of gravity, perturbations in the metric are generally described in terms of {\it two} potentials,
one of which is the Newtonian potential (apart from the deviations required to lead to MOND). 
The difference between the two potentials, also known as {\it gravitational slip}, can lead to changes in the growth of structure, large-scale gravitational lensing  and 
anisotropies in the CMB. For example, it is still unclear whether it is possible to completely match both 
the CMB data on large and small scales as well as the amplitude of mass fluctuations in galaxy surveys. If
one is to boost the small angle peaks of the predicted angular power spectrum of the CMB so that it can match
the data, one runs the risk of introducing large fluctuations on large scales due to the gravitational
slip. Furthermore this effect can suppress the amount of clustering on galactic, cluster and super cluster scales.
So as yet, the comparison between TeVeS, the CMB and large scale structure is not conclusive \cite{SkordisRev2009},
and all the more so for GEA and other modifications of gravity.

The gravitational slip may be the smoking gun for modified theories of gravity. There are, by now, a few suggestions
on how it may be detected. The idea is simple: Different cosmological data probe different combinations of
the two gravitational potentials and by combining them, it may be possible to tease out evidence for the slip.
So, for example, a galaxy catalogue will be a measurement of the density contrast of the Universe and will
be directly related to one of the potentials, whereas a map of large scale flows should probe the other potential. 
\begin{figure*}[htb]
\center
\epsfig{figure=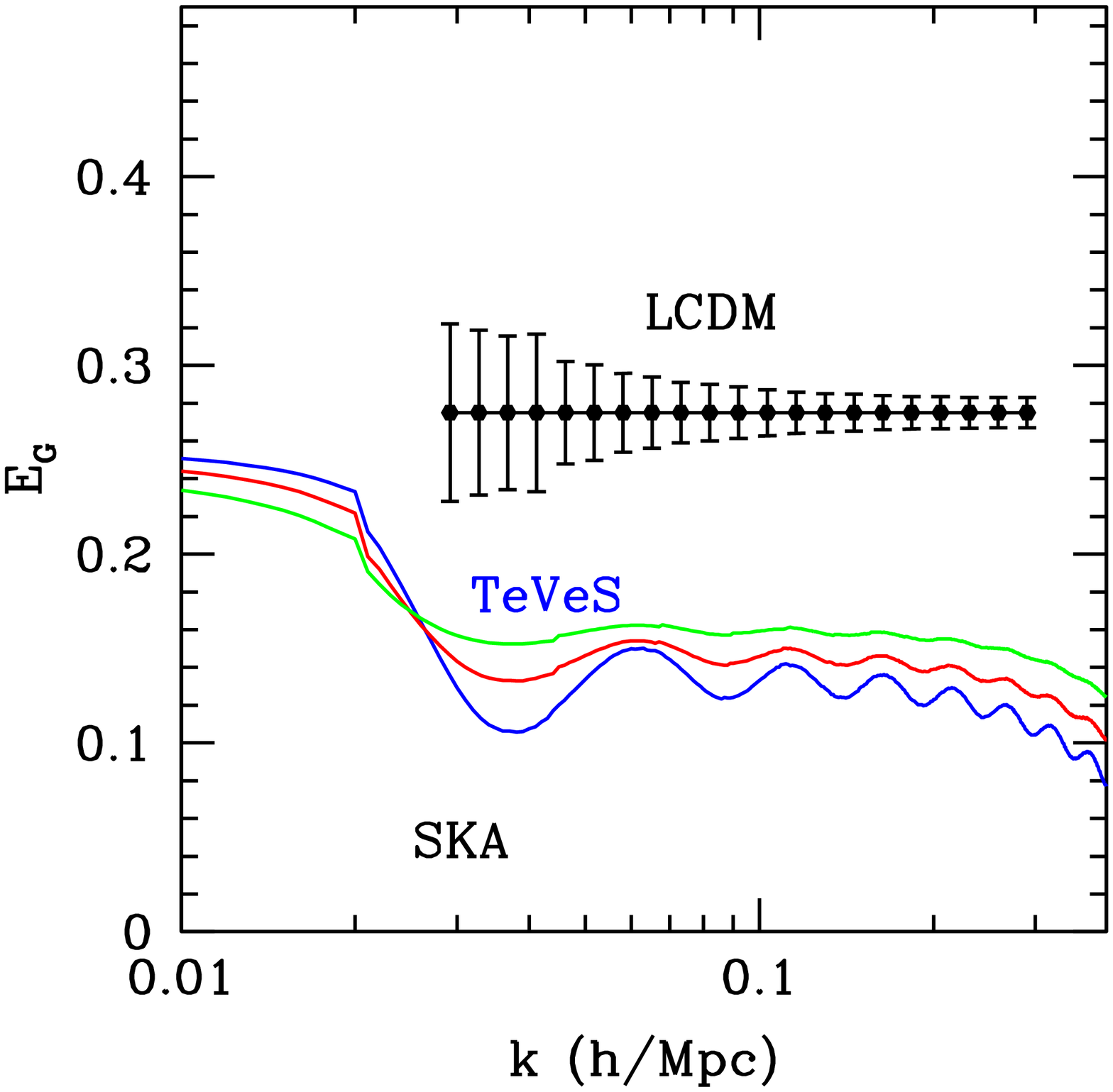,width=7cm}
\epsfig{figure=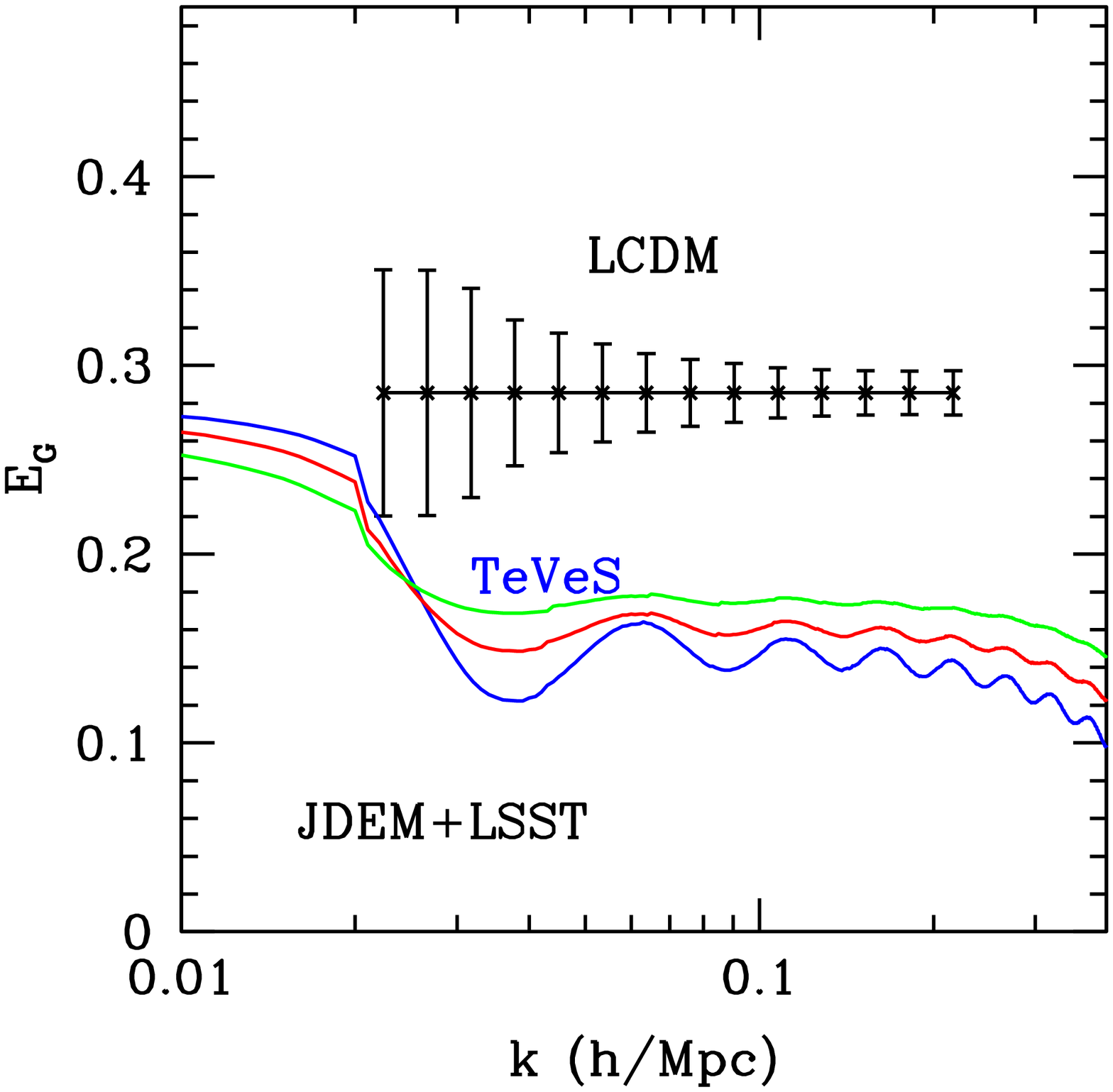,width=7cm}
\vspace{-10pt}
\caption{ 
The normalized cross correlation spectrum, $E_G$, between density and weak lensing on large scales,
as a function of wave number, $k$. The points and errors bars are a forecast of what would be expected in
a Universe with dark matter, a cosmological constant and Einstein gravity (LCDM) as measured by the Square Kilometre Array (SKA) or by a combination of the Large Synoptic Survey Telescope (LSST) and a possible version of the Joint Dark Energy Mission (JDEM) \cite{Zhang2007}. The
coloured lines are variants of the TeVeS model of modified gravity. The two different classes of theories are
clearly distinguishable (Figure courtesy of P.~Zhang, drawn from \cite{Zhang2007}).
}\label{fig3}
\end{figure*}
Measurements
of weak lensing will depend on the sum of the two potentials as will observations of the CMB. By cross-correlating
maps of the CMB with galaxy surveys, or alternatively matching maps of weak lensing with peculiar velocity
flows it should be possible to search for gravitational slip, and if found, it will give us information on about the most relevant
modifications to theories of gravity\cite{Lue2004,Zhang2007,Jain2008}.
Relativistic theories of modified gravity can be used to calculate the effects of gravitational lensing and can be tested
with the many well-measured gravitational lenses. A notable example is  the Bullet cluster where the baryonic mass, which is usually relatively well traced by hot x-ray emitting gas,
is severely misaligned with the sources of gravitational lensing as inferred from the distorted images of  background galaxies \cite{Clowe2007}. 
A dark matter explanation for the Bullet cluster is that the lensing is centered on two localized accumulations of dark matter, 
the halos of two colliding clusters that have passed through each other,
and the hot cluster gas is interacting as a result of that collision. 

It would appear difficult to reconstruct such a configuration merely by modifying gravity, but two features of such 
theories prevent such a simple assessment.
The first is the presence of the extra dynamical fields. 
Because structure is seeded in these models by fluctuations in these extra fields, unsourced by any baryonic fluctuations,
these fields clearly are capable of supporting fluctuations that are a source of gravity independent of the baryons.  
In principle, these gravity-field seeds may evolve into dark-field concentrations 
that interact only gravitationally,
become separated from their associated baryons in a collision, 
and source gravitational lensing as seen in the bullet cluster.

The second feature is that these theories  do not satisfy Birkhoff's theorem. 
As a result, not only is the gravitational field due to a localized concentration of matter that is not unique
(and may depend on the history that led to its assembly) but the environment can play a major 
role in the interactions between two systems. 
For example, in inferring the dynamics of galaxies within a cluster it becomes necessary to 
include the effects from the rest of the cluster, from neighbouring clusters and from any enveloping super-cluster \cite{Wu2007,Dai2008}. 
Yet, these theories should be predictive and there have been attempts to study lensing properties of specific galaxies and galaxy clusters  in TeVeS and GEA, with
mixed results \cite{Natarajan2008,Feix2008}. Without including the effect of extra degrees of freedom or the environment, it was found that, in TeVeS, the
Bullet cluster would have to be surrounded by massive neutrinos, with a mass of approximately $2$ electron volts \cite{Angus2007}.
Even with the inclusion of the effects of the extra degrees of freedom, 
it would be necessary to have some form of
dark matter, which could be in the form of neutrinos.   
This is all hardly surprising because, as discussed above, MOND requires clusters to have some dark matter.  
For GEA theories it is possible in principle to reconstruct the geometry 
and gravitational field of a lens such as the Bullet cluster, without any extra matter but with a substantial contribution from the extra degrees of freedom \cite{Dai2009}. 
The question of whether the dynamical evolution of the dark field perturbations leads
naturally to such large dark field halos at large time scales remains outstanding.

Although there is nothing intrinsically inconsistent with having the new fields that mediate the modifications of gravity
envisioned in MOND act as dark seeds of structure or dark concentrations of gravitational lensing, this necessity
 detracts from the cleanliness of the original MOND vision: What you see is apparently {\em not} what you
get, even in MOND. It is therefore much harder to make testable predictions for modified theories of gravity than 
was already thought and thus far, these relativistic extensions of MOND  remain viable solutions to the problem of missing mass.

\section*{Dark Energy and future measurements}
The problem of missing gravity has been at the forefront of cosmology for many decades. 
From the moment a proposal was put forward to solve the missing mass problem of galaxies with modified gravity, 
it was realized that there could be cosmological implications: The acceleration scale, $a_0\simeq10^{-8}cm/s^2$, 
which characterizes the transition between Newtonian and non-Newtonian gravity in MOND, 
is of the same order of magnitude as $cH_0$, 
where $c$ is the speed of light and $H_0$ is the expansion rate of the Universe today.

Furthermore, the recent discovery that the Universe may be accelerating (and not decelerating as GR predicts if the
energy content of the Universe is dominated by pressureless, non-relativistic matter) 
has been taken to imply the existence of an additional dark component of the Universe. 
Dubbed {\it dark energy} it is gravitationally repulsive and can drive the expansion at late times. 
It may be possible that the accelerated expansion of the Universe is instead due to modifications to  Einstein's theory of gravity.  

Some modified gravity theories seek to explain only the accelerated expansion and not the missing mass problem of galaxies and clusters.  DGP (Dvali-Gabadadze-Porrati) \cite{Deffayet2001} and (generic) $f(R)$ theories \cite{Carroll2003} are two that have received widespread attention.
However, as for unified models of the dark sector, where an extra component
of the energy density of the Universe can be {\it both} dark matter and dark energy, it is natural to consider models where modifications of gravity can give rise to both the breakdown of Newtonian gravity on galactic scales and to the acceleration of cosmic expansion,
as in TeVeS or GEA  \cite{Bourliot2007,Ratra2006,Zlosnik2007,Zhao2007,Blanchet2008}. 

As argued in the previous section, future surveys may be able to distinguish between 
theories of modified gravity and GR with dark matter and dark energy \cite{Zhang2007}. 
A survey of galaxies (such as those that will be done by the Joint Dark Energy Mission \cite{JDEM} or the 
Square Kilometre Array \cite{SKA}) combined
with a measurement of lensing on large scales (such the one proposed by the Joint Dark Energy Mission 
or the Large Scale Synpotic Survey \cite{LSST}), should be able to clearly  pick out the signature of gravitational slip of a theory such as TeVeS.
Furthermore, this would be on scales for which many of the issues that complicate predictions in the case
of quasi-isolated systems such as clusters, would not come into play.

{\underline {Acknowledgements:}} We thank Stacy McGaugh and Peng-Jie Zhang for kindly supplying us
with figures. We thank Constantinos Skordis and Tom Zlosnik for discussions. PGF is funded by STFC and
thanks the Beecroft Institute for support. GDS is funded by US-DOE, NASA and FXQI and thanks the Beecroft
Institute for hospitality.

\end{document}